\documentclass[pra,showpacs,reprint,superscriptaddress]{revtex4-1}

\usepackage{graphicx}
\usepackage{bm}
\usepackage{amsmath}
\usepackage{amsfonts}
\usepackage{color}

\newcommand{\beq}{\begin{equation}}
\newcommand{\eeq}{\end{equation}}
\newcommand{\beqa}{\begin{eqnarray}}
\newcommand{\eeqa}{\end{eqnarray}}

\begin{document}


\title{Single-qubit gates in two steps with rotation axes in a single plane}
 \author{Yun-Pil Shim}
 \altaffiliation{Current address: Laboratory for Physical Sciences, College Park, Maryland 20740, USA}
 \affiliation{Department of Physics, University of Wisconsin-Madison, Madison, Wisconsin 53706, USA}
 \author{Jianjia Fei}
 \affiliation{Department of Physics, University of Wisconsin-Madison, Madison, Wisconsin 53706, USA}
 \author{Sangchul Oh}
 \affiliation{Department of Physics, University at Buffalo, State University of New York, Buffalo, New York 14260, USA}
 \author{Xuedong Hu}
 \affiliation{Department of Physics, University at Buffalo, State University of New York, Buffalo, New York 14260, USA}
 \author{Mark Friesen}
 \affiliation{Department of Physics, University of Wisconsin-Madison, Madison, Wisconsin 53706, USA}
 \date{\today}

\begin{abstract}
Any single-qubit unitary operation or quantum gate can be considered a rotation. 
Typical experimental implementations of single-qubit gates involve two or three fixed rotation axes, and up to three rotation steps. 
Here we show that, if the rotation axes can be tuned arbitrarily in a fixed plane, then two rotation steps are sufficient for implementing a single-qubit gate, and one rotation step is sufficient for implementing a state transformation. 
The results are relevant for ``exchange-only" logical qubits encoded in three-spin blocks, which are important for universal quantum computation in decoherence free subsystems and subspaces.
\end{abstract}

\maketitle

\section{Introduction}
In the quantum circuit model \cite{deutsch_89}, a universal quantum computer requires an entangling two-qubit gate such as CNOT, and a set of single-qubit gates \cite{divincenzo_pra95}.
Although a finite set of quantum gates is sufficient for universality, fault-tolerant circuits require a large number of gates, even for very simple operations \cite{nielsen_chuang_book}.
It is therefore important to be able to perform gate operations as efficiently as possible, in order to minimize the effects of decoherence or gating errors.

Single-qubit operations can be viewed as rotations of the state vector of a logical qubit on a unit Bloch sphere \cite{nielsen_chuang_book}. 
The most efficient method for rotating a spin qubit would be to apply a magnetic field along the desired axis of rotation; 
however this is not practical for most physical implementations.
For example, to independently control an array of electron spin qubits in quantum dots \cite{loss_divincenzo_pra1998} would require an array of tunable micromagnets, which is experimentally challenging.

A more common approach is to provide two or three fixed, orthogonal rotation axes. 
This enables arbitrary rotations in three or fewer steps, for example, by using the Euler angle construction. 
Many qubit implementations employ this strategy.  
For single-spin qubits, this could involve tunable, external magnetic fields oriented along two orthogonal axes, although this could still be challenging in a scalable architecture.
For logical qubits composed of two or more physical spins, it is possible for different physical coupling mechanisms to control the different rotation axes on the Bloch sphere. 
For example, singlet-triplet logical qubits formed in double quantum dots use local magnetic field gradients to generate rotations about one axis, and exchange interactions to generate rotations about an orthogonal axis \cite{Petta2005}. 

\begin{figure}
  \includegraphics[width=\linewidth]{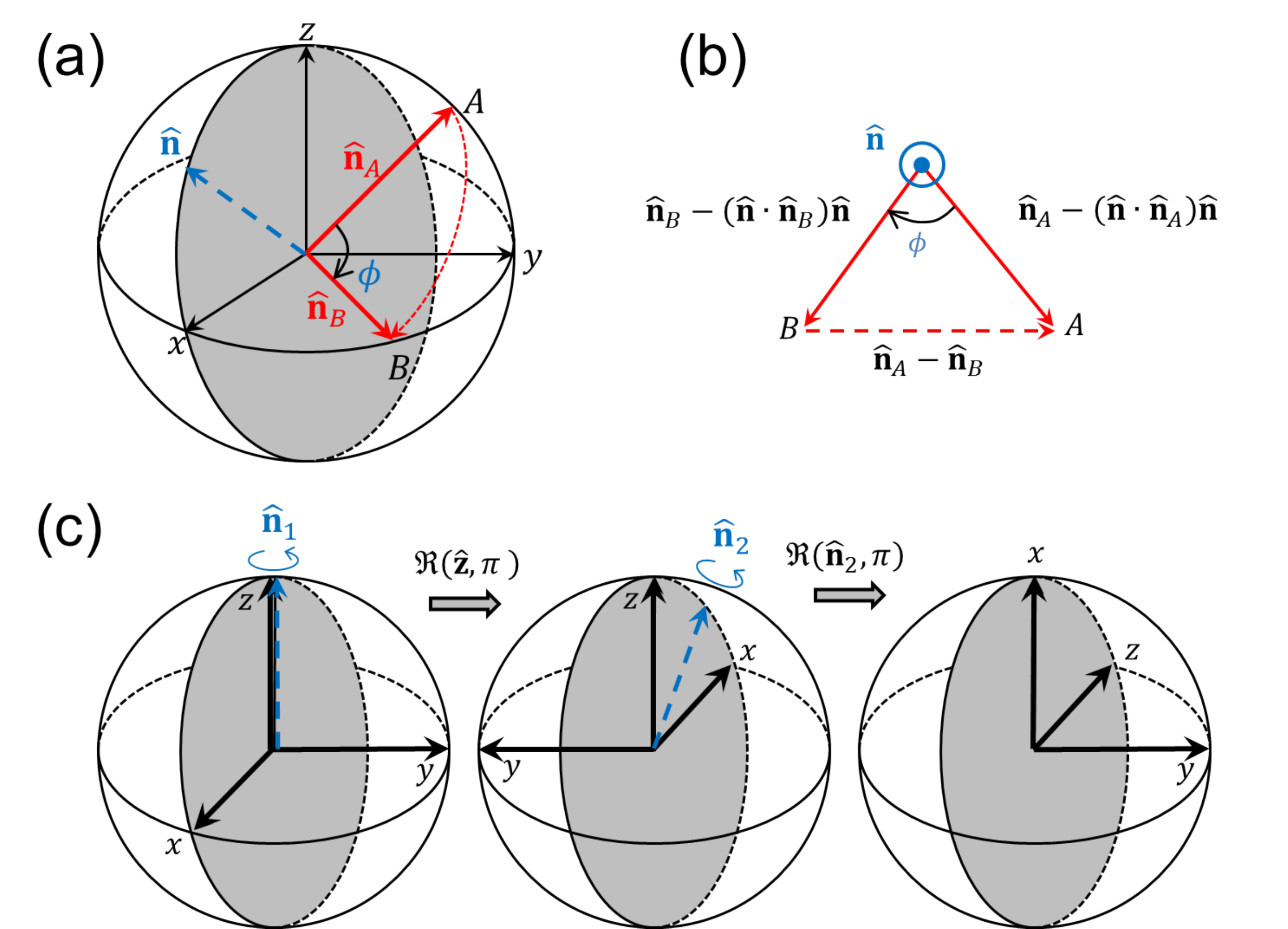}\\
  \caption{
  (Color online) Single-plane rotation method:  the shaded regions indicate the $xz$ plane of the Bloch sphere, and dashed blue arrows indicate the rotation axes.
  	(a) An arbitrary transformation, from the state $|\psi_A\rangle$ along $\hat{\mathbf{n}}_A$, 
	to the state $|\psi_B\rangle$ along $\hat{\mathbf{n}}_B$, via a single rotation 
           about the axis $\hat{\mathbf{n}}$. 
           In this example, $\hat{\mathbf{n}}_A$=$(0,1,1)/\sqrt{2}$ 
           and $\hat{\mathbf{n}}_B$=$(1,1,0)/\sqrt{2}$. 
           As described in the main text, we obtain the rotation axis
           $\hat{\mathbf{n}}$=$(1,0,1)/\sqrt{2}$, and rotation angle $\phi=-\cos^{-1} (1/3)$. 
           (c) An arbitrary single qubit rotation, performed in two steps. 
           In this example, $\mathcal{R}(\hat{\mathbf{y}},3\pi/2)$=$\mathcal{R}(\hat{\mathbf{n}}_2,\pi) \mathcal{R}(\hat{\mathbf{z}},\pi)$ 
           with $\hat{\mathbf{n}}_2$=$(-1,0,1)/\sqrt{2}$.
  }
  \label{fig:rotations}
\end{figure}

In this paper, we are concerned with intermediate situations, between these two extremes, where rotations can be performed about an arbitrary axis constrained to lie in a single plane. 
We consider the example of an exchange-only logical qubit in a triple quantum dot  \cite{divincenzo_bacon_nature2000,fong_wandzura_qic2011}, which is the main motivation for our work. 
Exchange-only qubits are of interest because they are formed in decoherence free subspaces and subsystems \cite{zanardi_rassetti_prl1997,duan_guo_pra1998,lidar_chuang_prl1998},
and they can be operated using only fast exchange interactions  \cite{bacon_kempe_prl2000,kempe_bacon_pra2001,kempe_bacon_qic2001}. 
A similar, single-plane rotation scheme is also possible for chirality-based logical qubits in a triple dot \cite{hsieh_hawrylak_prb2010}.   
In fact, for any system where the effective magnetic fields along two orthogonal axes can be tuned arbitrarily and simultaneously, the single-plane rotation condition is satisfied.

We will build upon the known result \cite{divincenzo_bacon_nature2000} that the internal couplings between the physical spins in an exchange-only qubit can be used to generate a continuous set of rotations in the $xz$ plane of the logical qubit.
We will show that such single-plane rotations reduce the total number of steps required for single-qubit operations. 
For example, a state transformation, which maps a specific initial state onto a specific final state, can be accomplished in one step, while an arbitrary single-qubit rotation can be accomplished in two steps. 
We provide constructive proofs for both of these problems.

\section{Single-step state transformation}
We first prove that a given initial state $|\Psi_A\rangle$ can be transformed to a specified state $|\Psi_B\rangle$, by a single rotation about an axis in the $xz$ plane. 
The Bloch sphere geometry is shown in Fig.~\ref{fig:rotations}(a).
Specifically, we want to determine the rotation axis $\hat{\mathbf{n}}$ and the rotation angle $\phi$ that satisfy
\begin{equation}
\mathcal{R}(\hat{\mathbf{n}},\phi) |\Psi_A\rangle = |\Psi_B\rangle ~,
\end{equation}
where the rotation operator $\mathcal{R}(\hat{\mathbf{n}},\phi)$=$\exp(-i\bm{\sigma}\cdot\hat{\mathbf{n}}\phi/2)$ is defined in terms of the Pauli spin matrix vector $\bm{\sigma}$.
The qubit state vectors $\hat{\mathbf{n}}_A$ and $\hat{\mathbf{n}}_B$ are also pictured on the Bloch sphere of Fig.~\ref{fig:rotations}(a).
If the rotation axis is allowed to point in any direction (not just the $xz$ plane), then we could choose 
$\hat{\mathbf{n}} \propto  ( \hat{\mathbf{n}}_A + \hat{\mathbf{n}}_B )$ with $\phi=\pi$.   
We now show that the desired result can be achieved, even when $\hat{\mathbf{n}}$ is confined to the $xz$ plane. 

To perform a state transformation in a single step, it is clear that the rotation axis must be equidistant from both $\hat{\mathbf{n}}_A$ and $\hat{\mathbf{n}}_B$. 
This constraint defines a plane, given by $\hat{\mathbf{n}} \cdot ( \hat{\mathbf{n}}_A - \hat{\mathbf{n}}_B ) = 0$. On the other hand, we require the rotation axis to lie in the $xz$ plane, which is defined by $\hat{\mathbf{n}} \cdot \hat{\mathbf{y}} = 0$. 
The intersection of the two planes is given by
\begin{equation}\label{eq:n_state_tf}
\hat{\mathbf{n}} = \frac{\hat{\mathbf{y}} \times ( \hat{\mathbf{n}}_A - \hat{\mathbf{n}}_B )}
                        { \left| \hat{\mathbf{y}} \times ( \hat{\mathbf{n}}_A - \hat{\mathbf{n}}_B ) \right|} ~.
\end{equation}
Figure~\ref{fig:rotations}(b) shows the projection of $\hat{\mathbf{n}}_A$ and $\hat{\mathbf{n}}_B$ in the plane perpendicular to $\hat{\mathbf{n}}$.
The inscribed angle is the angle of rotation, $\phi$. 
Since the length of all three sides of the triangle are known, as indicated in the figure, we can obtain
\begin{equation}
\cos\phi=1-|\hat{\mathbf{n}}_A-\hat{\mathbf{n}}_B|^2/2L^2, \label{eq:cosphi}
\end{equation}
where $L=|\hat{\mathbf{n}}_A-(\hat{\mathbf{n}}\cdot\hat{\mathbf{n}}_A)\hat{\mathbf{n}}| = |\hat{\mathbf{n}}_B-(\hat{\mathbf{n}}\cdot\hat{\mathbf{n}}_B)\hat{\mathbf{n}}|$. 
The sign of $\phi$ is given by $\text{sgn}[(\hat{\mathbf{n}}_A \times \hat{\mathbf{n}}_B)\cdot\hat{\mathbf{n}}]$.
When the two planes are equivalent, we cannot use Eq.~(\ref{eq:n_state_tf}). 
However in this special case,  $\hat{\mathbf{n}}_A + \hat{\mathbf{n}}_B$ lies in the $xz$ plane, and we can simply choose $\hat{\mathbf{n}} \propto ( \hat{\mathbf{n}}_A + \hat{\mathbf{n}}_B )$ with $\phi=\pi$.  

Figure~\ref{fig:rotations}(a) shows an example of the state transformation procedure for the case $\hat{\mathbf{n}}_A$=$(0,\frac{\sqrt{2}}{2},\frac{\sqrt{2}}{2})$ and $\hat{\mathbf{n}}_B$=$(\frac{\sqrt{2}}{2},\frac{\sqrt{2}}{2},0)$. 
From Eqs.~(\ref{eq:n_state_tf}) and (\ref{eq:cosphi}), we obtain the rotation axis $\hat{\mathbf{n}}$=$(\frac{\sqrt{2}}{2},0,\frac{\sqrt{2}}{2})$ and the rotation angle $\phi=-\cos^{-1}(1/3)\simeq -70.53^{\circ}$.

\section{Two-step qubit rotations}
\begin{figure}
  \includegraphics[width=\linewidth]{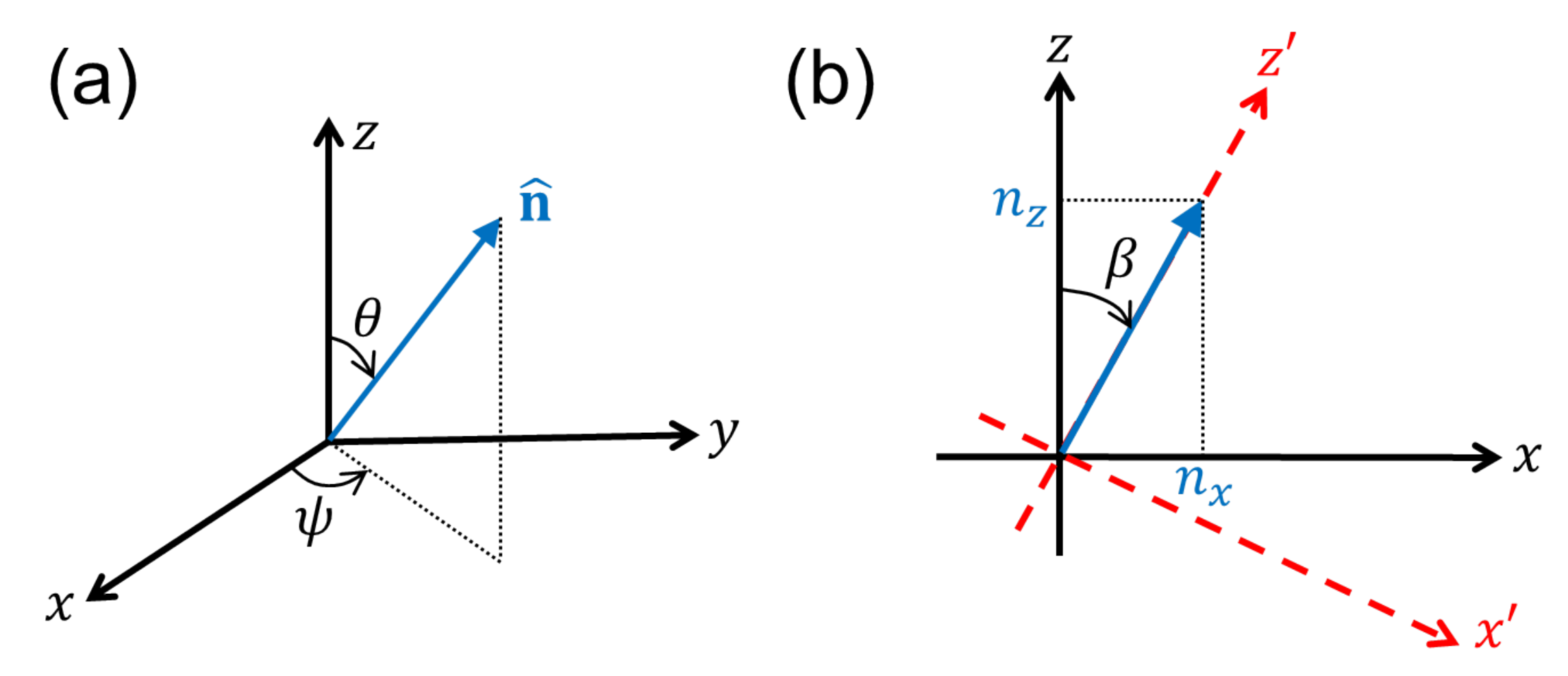}\\
  \caption{
  (Color online) Definition of the angular variables.
  (a) The logical rotation axis $\hat{\mathbf{n}}$, with the polar angle $\theta$ and the azimuthal angle $\psi$. 
  (b) The projection of $\hat{\mathbf{n}}$ onto the $xz$ plane.  
  We define a new coordinate system with axes $\hat{\mathbf{x}}'$ and $\hat{\mathbf{z}}'$, such that $n_x'=0$, by rotating the $xz$ plane around the $y$ axis by angle $\beta$.}
  \label{fig:new_axes}
\end{figure}

We now provide a constructive proof that any single-qubit gate (up to a global phase) can be generated in two rotation steps, when the rotation axes point in an arbitrary direction in a single plane. 
Specifically, we want to solve for the rotation parameters defined by
\begin{equation}\label{eq:twostep}
\mathcal{R}(\hat{\mathbf{n}},\phi) = 
e^{i\eta} \mathcal{R}_2(\hat{\mathbf{n}}_2,\phi_2)  \mathcal{R}_1(\hat{\mathbf{n}}_1,\phi_1) ~,
\end{equation}
where the rotation axis $\hat{\mathbf{n}}$ can point anywhere in the Bloch sphere, but the individual rotation axes $\hat{\mathbf{n}}_1$ and $\hat{\mathbf{n}}_2$ lie in the $xz$ plane.
It is convenient to work with angular coordinates defined by
\begin{gather}
\hat{\mathbf{n}} = \left( \sin\theta \cos\psi, \sin\theta \sin\psi, \cos\theta \right) ~, \\
\hat{\mathbf{n}}_1 = \left( \sin\theta_1 , 0, \cos\theta_1 \right) ~, \\
\hat{\mathbf{n}}_2 = \left( \sin\theta_2 , 0, \cos\theta_2 \right) ~,
\end{gather}
and illustrated in Fig.~\ref{fig:new_axes}(a).
Since an arbitrary rotation is characterized by three parameters ($\theta,\psi,\phi$), while the right-hand-side of Eq.~(\ref{eq:twostep}) involves five parameters ($\eta,\theta_1,\phi_1,\theta_2,\phi_2$), the equation is clearly under-constrained; many solutions exist, any of which suits our needs.

We now demonstrate that at least one solution exists by providing an explicit, analytical construction.
We first simplify the problem by transforming to a new set of coordinate axes defined by the projection of the logical rotation axis $\hat{\mathbf{n}}$ onto the $xz$ plane.
The normalized projection becomes our new $\hat{\mathbf{z}}$ axis, as shown in Fig.~\ref{fig:new_axes}(b).
The rotation angle for the transformation, $\beta$, is given by
\beq
\label{eq:beta}
\cos\beta = \frac{n_z}{\sqrt{n_x^2+n_z^2}}~, \quad
\sin\beta = \frac{n_x}{\sqrt{n_x^2+n_z^2}}~. 
\eeq 
where $n_x$ and $n_z$ are the components of $\hat{\mathbf{n}}$ in the original coordinate system.
Using primed variables to indicate the new coordinate system, we have 
\begin{equation}
\hat{\mathbf{n}} = \sin\theta' \hat{\mathbf{y}} + \cos\theta' \hat{\mathbf{z}'} ~.
\end{equation}
where
\beq
\label{eq:thetap}
\sin\theta' = \sin\theta\sin\psi ~.
\eeq
Note that $\hat{\mathbf{y}}$ is the same in both coordinate systems.

We now solve Eq.~(\ref{eq:twostep}) in the primed coordinate system.
Without loss of generality, we will restrict the two rotation axes in the range $\theta_1',\theta_2' \in [0,\pi)$ and the rotation angles in the range $\phi_1,\phi_2 \in [0,2\pi)$, since a rotation with $\theta_i' \geq \pi$ by an angle $\phi_i$ is equivalent to the rotation with $\theta_i' - \pi$ by an angle $2\pi - \phi_i$.
We can take advantage of the under-constrained nature of the problem by making the convenient choice $\phi_2=\pi$. 
(See Appendix~\ref{sec:appendixB} for details.)
We then find
\begin{gather}
\theta'_2 = 0 ~, \label{eq:thetap2} \\
k\cos\theta'\sin\frac{\phi}{2} = \cos\frac{\phi_1}{2} ~,\label{eq:phi1} \\
k\sin\theta'\tan\frac{\phi}{2} = -\tan\theta'_1 ~, \label{eq:theta1p}
\end{gather}
where $k=\text{sgn} [n_y]=\text{sgn} [\sin\theta \sin\psi]$.
Here, $\theta'_2$, $\phi_1$, and $\theta'_1$ are determined from Eqs.~(\ref{eq:thetap2}), (\ref{eq:phi1}), and (\ref{eq:theta1p}), respectively.  
These quantities are related to the original coordinate system through $\theta_1$=$\theta'_1+\beta$ and $\theta_2$=$\theta'_2+\beta$.

As noted above, Eqs.~(\ref{eq:thetap2})-(\ref{eq:theta1p}) do not represent unique solutions to Eq.~(\ref{eq:twostep}). 
For example, we may obtain an alternative solution with the choice $\phi_1=\pi$. 
After a similar analysis, we then obtain
\begin{gather}
\theta'_1=0~, \\
-k\cos\theta'\sin\frac{\phi}{2} = \cos\frac{\phi_2}{2}~, \\
k\sin\theta'\tan\frac{\phi}{2} = \tan\theta'_2~. 
\end{gather} 

To provide a practical comparison, we consider a conventional (Euler) method vs.\ the single-plane method, for the specific case of rotations around the $\hat{\mathbf{y}}$ axis.
For the Euler method, if we have two fixed axes of rotation ($\hat{\mathbf{x}}$ and $\hat{\mathbf{z}}$), the result corresponds to a three-step procedure given by $\mathcal{R}(\hat{\mathbf{y}},\phi)=\mathcal{R}(\hat{\mathbf{z}},\pi/2)\mathcal{R}(\hat{\mathbf{x}},\phi)\mathcal{R}(\hat{\mathbf{z}},-\pi/2)$. 
In contrast, the single-plane method is accomplished in just two steps, with $\mathcal{R}(\hat{\mathbf{y}},\phi)=\mathcal{R}(\hat{\mathbf{n}}_2,\pi) \mathcal{R}(\hat{\mathbf{z}},\pi) \label{eq:y_rot_2}$
where $\hat{\mathbf{n}}_2$=$(\sin\frac{\phi}{2},0,\cos\frac{\phi}{2})$. 
Figure~\ref{fig:rotations}(c) shows an explicit construction of a $3\pi/2$ rotation about the $\hat{\mathbf{y}}$ axis, employing two rotations around axes in the $xz$ plane.

\section{Rotations for exchange-only qubits}
We now consider a concrete physical example:  a logical exchange-only qubit encoded in a three-spin block, for which  
all qubit operations are implemented using electrically tunable exchange couplings between the constituent spins, without requiring a magnetic field \cite{divincenzo_bacon_nature2000}.
The effective Hamiltonian for the spin system is given by
\begin{equation}\label{eq:Hexch}
\widehat{H} = J_{12} \mathbf{S}_1\cdot\mathbf{S}_2 +  J_{23} \mathbf{S}_2\cdot\mathbf{S}_3 +  J_{31} \mathbf{S}_3\cdot\mathbf{S}_1 ~,
\end{equation}
where the exchange coupling parameters $J_{12}$, $J_{23}$, and $J_{31}$ are typically non-negative.
The total spin $S_{\mathrm{tot}}$ and its $z$ component $S_{\mathrm{tot}}^z$ are good quantum numbers since they commute with the Hamiltonian; we will use them to label the energy eigenstates.
We are specifically interested in the states with $S_{\mathrm{tot}}$=1/2 and $S_{\mathrm{tot}}^z$=$\pm1/2$, which are both two-fold degenerate.
We specify these states as $ \{ |S_{\mathrm{tot}},S_{\mathrm{tot}}^z;l \rangle \}$, adopting the label $l$=0,1 for the degenerate states.
With these definitions, we can encode the qubit in a decoherence free subsystem as follows \cite{fong_wandzura_qic2011}: 
\begin{eqnarray}
|0\rangle_L &=& a \left| \frac{1}{2},\frac{1}{2};0 \right\rangle + b \left|\frac{1}{2},-\frac{1}{2};0 \right\rangle ~,  
\label{eq:subsystem0} \\
|1\rangle_L &=& a \left|\frac{1}{2},\frac{1}{2};1 \right\rangle + b \left|\frac{1}{2},-\frac{1}{2};1 \right\rangle   ~,
\label{eq:subsystem1} \
\end{eqnarray}
where $|\frac{1}{2},\frac{1}{2};0 \rangle$=$|S\rangle_{12}\otimes|\uparrow\rangle_3$, $|\frac{1}{2},\frac{1}{2};1\rangle$=$-\sqrt{1/3}|T_0\rangle_{12}\otimes|\uparrow\rangle_3$+$\sqrt{2/3}|T_+\rangle_{12}\otimes|\downarrow\rangle_3$, and $|\frac{1}{2},-\frac{1}{2};0 \rangle$ and $|\frac{1}{2},-\frac{1}{2};1 \rangle$ are defined as spin-flipped states of $|\frac{1}{2},\frac{1}{2};0 \rangle$ and $|\frac{1}{2},\frac{1}{2};1 \rangle$, respectively.

In the logical qubit space, the Hamiltonian becomes \cite{weinstein_hellberg_pra2005}
\begin{eqnarray}\label{eq:Hpauli}
\widehat{H} &=& -\frac{J_{12}+J_{23}+J_{31}}{4} \openone 
    + \frac{\sqrt{3}\left( J_{23}-J_{31} \right)}{4} \sigma_x  \nonumber\\
 && + \frac{ -2J_{12}+J_{23}+J_{31}}{4} \sigma_z~.  
\end{eqnarray}
The unitary operator $\widehat{U}(t)$=$\exp(-i\widehat{H}t/\hbar)$ rotates the logical qubit around an axis $\hat{\mathbf{n}}$ in the $xz$ plane by an angle $\phi$, given by
\begin{gather}
\hat{\mathbf{n}} = \frac{1}{2J} 
  \left( \sqrt{3}\left( J_{23}-J_{31} \right), 0 , 
         -2J_{12}+J_{23}+J_{31} \right)  , \label{eq:n_exchange}\\ 
\phi = J t /\hbar~, \label{eq:phi_exchange}
\end{gather}
where $J$=$\sqrt{J_{12}^2+J_{23}^2+J_{31}^2-J_{12}J_{23}-J_{23}J_{31}-J_{31}J_{12}}$.
The decoherence free subspace considered in Ref.~\cite{divincenzo_bacon_nature2000} corresponds to the special case of $a$=1 and $b$=0 in Eqs.~(\ref{eq:subsystem0}) and (\ref{eq:subsystem1}).

\begin{figure}
  \includegraphics[width=\linewidth]{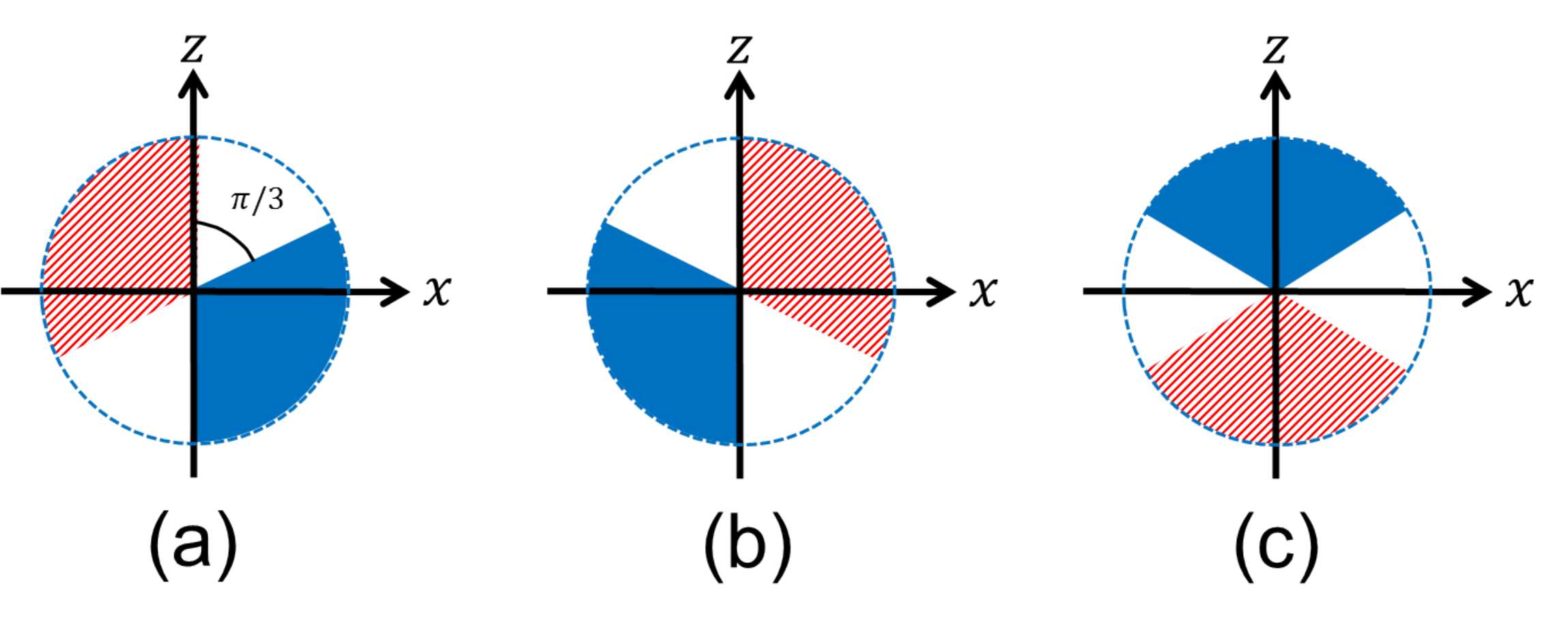}\\
  \caption{(Color online) Regions of the $xz$ plane where viable rotation axes are found, when one of the three exchange couplings is set to zero: 
  (a) $J_{31}=0$; (b) $J_{23}=0$; (c) $J_{12}=0$. 
  Solid blue regions represent the rotation axes $\hat{\mathbf{n}}$, consistent with Eq.~(\ref{eq:n_exchange}), assuming non-negative exchange couplings. 
  Striped, red regions correspond to rotations around the $-\hat{\mathbf{n}}$ axis by angle $\phi$, which is equivalent to rotations around $\hat{\mathbf{n}}$ by angle $2\pi-\phi$. 
  }
  \label{fig:rotations_with_two_Js}
\end{figure}

Equation (\ref{eq:Hexch}) suggests a ring-like coupling configuration for the quantum dots.  
Such configurations have been achieved in the laboratory \cite{gaudreau_studenikin_prl2006,amaha_hatano_apl2009}; 
however full control of the couplings is challenging.  
When full control is possible, the qubit can be rotated along any axis in the $xz$ plane. 
Single-qubit operations then follow the procedures described above.

A more common experimental arrangement is the linear triple quantum dot geometry, with one electron per dot. In this configuration, great tunability can be attained in the exchange couplings \cite{schroer_greentree_prb2007,laird2010,amaha_hatano_prb2012,gaudreau_granger_natphys2012,Busl_granger_natnano2013,medford_preprint}.
However, since one of the exchange couplings in Eq.~(\ref{eq:Hexch}) is assumed to vanish, it is not possible to implement arbitrary rotations in the $xz$ plane; only 2/3 of the plane is covered.
For example, if $J_{31}=0$, rotations are limited to the range $\pi/3 \leq \theta \leq \pi$ and $0 \leq \theta \leq 2\pi/3$. 
Figure \ref{fig:rotations_with_two_Js} shows the viable rotation axes in the $xz$ plane when one of the exchange couplings is set to zero. 

Despite the fact that we cannot perform rotations over the entire $xz$ plane for the linear dot geometry, many gates of interest can still be implemented in two or fewer steps. 
Focusing on the configuration $J_{13}=0$, we see that rotations about the $\hat{\bf x}$ axis ($J_{23}=2J_{12}$) and the $\hat{\bf z}$ axis ($J_{23}=0$) can be accomplished in a single step.
Rotations around the $\hat{\bf y}$ axis by an angle $\phi$ can be accomplished in two steps, 
by using one of the two solutions 
\begin{equation}
\mathcal{R}(\hat{\mathbf{y}},\phi) 
  = \mathcal{R}(\hat{\mathbf{z}},\pi) \mathcal{R}(\hat{\mathbf{n}}_1,\pi) 
  = -\mathcal{R}(\hat{\mathbf{n}}_2,\pi) \mathcal{R}(\hat{\mathbf{z}},\pi) \label{eq:y_rot} ~,
\end{equation}
where $\hat{\mathbf{n}}_1$=$(\sin\frac{\phi}{2},0,-\cos\frac{\phi}{2})$ and $\hat{\mathbf{n}}_2$=$(\sin\frac{\phi}{2},0,\cos\frac{\phi}{2})$. 
Up to a global phase, the phase gate $S$=$[(1,0),(0,i)]$ and the $\pi/8$ gate $T$=$[(1,0),(0,\exp(i\pi/4))]$ can be implemented with single-step rotations around the $\hat{\bf z}$ axis.

The Hadamard gate $H$=$1/\sqrt{2} [(1,1),(1,-1)]$ corresponds to a rotation around an axis with $\theta=\pi/4$ and $\psi=0$, by an angle $\phi=\pi$. 
For the configuration with $J_{12}=0$, this can be implemented in a single step.
However, in the $J_{23}=0$ or $J_{31}=0$ configurations, it cannot be implemented in fewer than three steps. 
If many Hadamard gates are required for a given quantum circuit, this could present a bottleneck.
In this case, it might be preferable to change the logical qubit basis in Eqs.~(\ref{eq:subsystem0}) and (\ref{eq:subsystem1}), to one where qubits 2 and 3 are exchanged.
In the latter configuration, the Hadamard gate is accomplished in one step.

\begin{figure}
  \includegraphics[width=\linewidth]{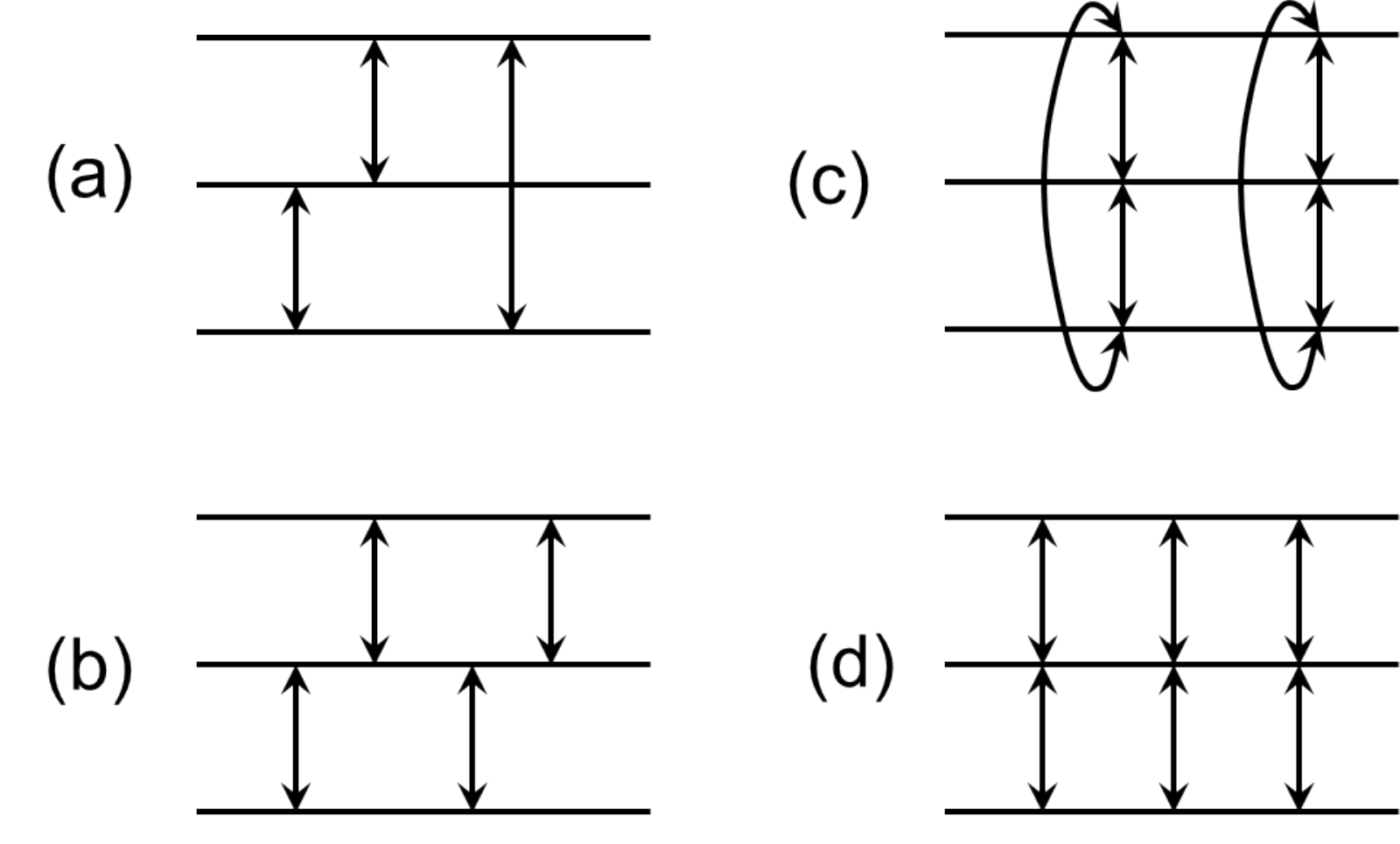}\\
  \caption{ 
  Logical qubit rotation schemes for exchange-only logical qubits.
  The horizonal lines depict the three physical spins comprising the logical qubit, and the vertical arrows indicate exchange couplings between two spins.
  The gating scheme of Ref.~\cite{divincenzo_bacon_nature2000} considers serial exchange couplings in (a) a ring geometry, or (b) a linear geometry.
  The single-plane rotation scheme described here considers simultaneous, parallel exchange couplings in (c) a ring geometry, or (d) a linear geometry.
  In the ring geometry, logical qubit gates require up to three steps in serial mode [(a)], and up to two steps in parallel mode [(c)].
  In the linear geometry, logical qubit gates require up to four steps in serial mode [(b)], and up to three steps in parallel mode [(d)].
  }
  \label{fig:comparison}
\end{figure}

We can compare our single-plane rotation method with the serial gating scheme for exchange-only qubits, which was described in \cite{divincenzo_bacon_nature2000}.
There it was shown that when only one exchange coupling ($J_{12}$, $J_{23}$, or $J_{31}$)  is allowed at a time, then a general logical qubit rotation requires three (four) steps for a ring (linear) geometry, as illustrated in Figs.~\ref{fig:comparison}(a) and (b).
The single-plane rotation method describe above can be viewed as a \emph{parallel} gating scheme, where multiple exchange couplings are implemented simultaneously, as illustrated in Figs.~\ref{fig:comparison}(c) and (d).
As we have shown, when two (three) exchange couplings are allowed simultaneously, then a logical qubit rotation requires three (two) steps.
Comparison with Figs.~\ref{fig:comparison}(a) and (b) shows that the single-plane method is always more effecient than the serial gating scheme.

\section{Conclusions}
We have shown that the ability to tune qubit rotation axes in a single, fixed plane enables  efficient, two-step implementations of single-qubit gates.
This should be contrasted with fixed-axis methods (e.g., Euler angles), that require up to three steps.
Our results can be applied directly to logical qubits in the decoherence free subspaces and subsystems of three-spin logical qubits, such as exchange-only qubits.

Our scheme can also be adapted to any qubit implementation with at least partial control over the rotation axes. 
For example, the effective field corresponding to rotations of a singlet-triplet logical qubit \cite{Petta2005} consists of a fixed $B_x$ component and a tunable, positive $B_z$ component.
The resulting effective rotation axis covers about half of the $xz$ plane \cite{wang_bishop_ncomm2012,kestner_wang_arxiv2013}, thus enabling efficient gating methods similar to those described here.

This work was supported by the DARPA QuEST program through a grant from AFOSR, by NSA/LPS through grants from ARO (W911NF-08-1-0482, W911NF-12-1-0607, and W911NF-09-1-0393), and by the DOD. The views and conclusions contained in this document are those of the authors and should not be interpreted as representing the official policies, either expressly or implied, of the US Government.

\begin{appendix}

\section{Rotation operators}

A rotation operator $\mathcal{R}(\hat{\mathbf{n}},\phi)$ is defined by its rotation axis $\hat{\mathbf{n}}$ and rotation angle $\phi$. The rotation axis is described by a polar angle $\theta$ in the range $[0,\pi]$ and an azimuthal angle $\psi$ in the range $[0,2\pi)$:
\beq
\hat{\mathbf{n}} = \left( \sin\theta \cos\psi, \sin\theta \sin\psi, \cos\theta \right)~.
\eeq
The rotation angle $\phi$ is also in the range $[0,2\pi)$.
In terms of Pauli operators, the rotation operator can be written as
\beqa
\mathcal{R}(\hat{\mathbf{n}},\phi) &=& \exp \left[ -i \frac{\bm{\sigma}\cdot\hat{\mathbf{n}}}{2} \phi \right] \nonumber\\
&=& \openone \cos\frac{\phi}{2} - i \bm{\sigma}\cdot\hat{\mathbf{n}} \sin\frac{\phi}{2} ~. 
\eeqa

Any rotation operators in a two-dimensional Hilbert space can be represented as
\beq
\mathcal{R} = a_0 \openone + i ( a_1 \sigma_x + a_2 \sigma_y + a_3 \sigma_z ) ~,
\eeq
where $a_{0,1,2,3}$ are all real.
In matrix form, 
\beq
\mathcal{R} =
\left( \begin{array}{cc} a_0 + i a_3  &  a_2 + i a_1 \\
                         -a_2 + i a_1 & a_0 - i a_3
       \end{array}
\right) ~.                       
\eeq
The unimodular condition requires that $a_0^2+a_1^2+a_2^2+a_3^2=1$.

The necessary and sufficient condition for two rotation operators $\mathcal{R}_1 = a_0 \openone + i ( a_1 \sigma_x + a_2 \sigma_y + a_3 \sigma_z )$ and
$\mathcal{R}_2 = b_0 \openone + i ( b_1 \sigma_x + b_2 \sigma_y + b_3 \sigma_z )$ to be identical, 
is that $a_i=b_i$ for $i=0,1,2,3$. This is obvious by comparing the matrix elements of the two rotation operators.

\section{Two-step implementations of single qubit gates}\label{sec:appendixB}

We want to identify two sequential rotation steps around axes in the $xz$ plane that produce an arbitrary single-qubit gate, up to a global phase:
\beq\label{eq:suppl_twostep}
\mathcal{R}(\hat{\mathbf{n}},\phi) = e^{i\eta} \mathcal{R}(\hat{\mathbf{n}}_2,\phi_2)  \mathcal{R}(\hat{\mathbf{n}}_1,\phi_1) .
\eeq
According to the constraints discussed in the main text,
$\hat{\mathbf{n}}$ can be any direction, while $\hat{\mathbf{n}}_1$ and $\hat{\mathbf{n}}_2$ must lie in the $xz$ plane:
\begin{gather}
\hat{\mathbf{n}} = \left( \sin\theta \cos\psi, \sin\theta \sin\psi, \cos\theta \right)~, \\
\hat{\mathbf{n}}_1 = \left( \sin\theta_1 , 0, \cos\theta_1 \right) ~, \\
\hat{\mathbf{n}}_2 = \left( \sin\theta_2 , 0, \cos\theta_2 \right) ~.
\end{gather}

We now transform the problem to a new set of (primed) coordinate axes, for which $\hat{\mathbf{n}}$ lies in the new $y'z'$ plane:
\beq
\hat{\mathbf{n}} = \sin\theta' \hat{\mathbf{y}}' + \cos\theta' \hat{\mathbf{z}}' ~.
\eeq
This can be achieved by rotating the original coordinate axes around $\hat{\mathbf{y}}$ axis. 
(See Fig.2 in the main text.)
Hence, the $\hat{\mathbf{y}}$ axis is unaffected, while the $\hat{\mathbf{x}}$ and $\hat{\mathbf{z}}$ axes become $\hat{\mathbf{x}}'$ and $\hat{\mathbf{z}}'$.
The rotation angle $\beta$ for this transformation is given by 
\beq
\cos\beta = \frac{n_z}{\sqrt{n_x^2+n_z^2}}~, \quad
\sin\beta = \frac{n_x}{\sqrt{n_x^2+n_z^2}}~. 
\eeq 
In the primed coordinate system, we easily find that $\psi'=\pi/2$ when $n_y > 0$, and $\psi'=3\pi/2$ when $n_y < 0$.
$\theta'$ is obtained from
\begin{gather}
\sin\theta' = \sin\theta\sin\psi ~. 
\end{gather}
Expanding both sides of Eq.~(\ref{eq:suppl_twostep}) in terms of Pauli operators and matching their coefficients, 
we obtain the following relations in the primed coordinate system:
\begin{eqnarray}
&&\cos\frac{\phi}{2} = e^{i\eta} \left[ \cos\frac{\phi_2}{2}\cos\frac{\phi_1}{2} 
                                      - \cos(\theta'_2-\theta'_1)\sin\frac{\phi_2}{2}\sin\frac{\phi_1}{2}
                                  \right]~, \nonumber \\ && \\                                  
&& 0 = e^{i\eta} \left[ \sin\theta'_1\cos\frac{\phi_2}{2}\sin\frac{\phi_1}{2} 
                                                        + \sin\theta'_2\sin\frac{\phi_2}{2}\cos\frac{\phi_1}{2} 
                                                   \right]~, \\
&& k \sin\theta'\sin\frac{\phi}{2} = e^{i\eta} \left[ -\sin(\theta'_2-\theta'_1) \sin\frac{\phi_2}{2}\sin\frac{\phi_1}{2} 
                                                   \right]~, \\
&& \cos\theta'\sin\frac{\phi}{2} = e^{i\eta} \left[ \cos\theta'_1\cos\frac{\phi_2}{2}\sin\frac{\phi_1}{2} \right. \nonumber\\
&&                           \hspace{1.1in} \left. + \cos\theta'_2\sin\frac{\phi_2}{2}\cos\frac{\phi_1}{2} \right] ~.
\end{eqnarray}

Here, we define $k=\text{sgn} [n_y]=\text{sgn} [\sin\theta \sin\psi]$.
Numerical results suggest that there will be a continuum of solutions for these equations. 
Here we set $\phi_2=\pi$, to simplify the equations and to enable an analytical solution. 
In this case, the equations reduce to
\begin{gather}
\cos\frac{\phi}{2} = - e^{i\eta} \cos(\theta'_2-\theta'_1)\sin\frac{\phi_1}{2} ~, \label{eq:suppl_cond1} \\
0 = e^{i\eta} \sin\theta'_2\cos\frac{\phi_1}{2} ~, \label{eq:suppl_cond2}\\
k \sin\theta'\sin\frac{\phi}{2} = - e^{i\eta}\sin(\theta'_2-\theta'_1)\sin\frac{\phi_1}{2} ~, \label{eq:suppl_cond3}\\
\cos\theta'\sin\frac{\phi}{2} = e^{i\eta} \cos\theta'_2\cos\frac{\phi_1}{2} ~.\label{eq:suppl_cond4}
\end{gather}
From Eq.~(\ref{eq:suppl_cond2}), we see that either $\sin\theta'_2=0$, or $\cos\frac{\phi_1}{2}=0$. 
To satisfy Eq.~(\ref{eq:suppl_cond4}), we must have $\sin\theta'_2=0$.
We therefore obtain $\theta'_2=0$. 

The three remaining equations are
\begin{gather}
\cos\frac{\phi}{2} = - e^{i\eta} \cos\theta'_1\sin\frac{\phi_1}{2} ~, \label{eq:suppl_c1} \\
k\sin\theta'\sin\frac{\phi}{2} = e^{i\eta}\sin\theta'_1\sin\frac{\phi_1}{2} ~,\label{eq:suppl_c2}\\
\cos\theta'\sin\frac{\phi}{2} = e^{i\eta}\cos\frac{\phi_1}{2} ~.\label{eq:suppl_c3}
\end{gather}
From Eq.~(\ref{eq:suppl_c2}), we see that $e^{i\eta}=k$, since the sine functions are all positive for the range of angles $\theta',\theta_1' \in [0,\pi]$ and $\phi,\phi_1 \in [0,2\pi)$. 
Hence,
\begin{gather}
k\cos\frac{\phi}{2} = -\cos\theta'_1\sin\frac{\phi_1}{2} ~, \label{eq:suppl_newc1} \\
\sin\theta'\sin\frac{\phi}{2} = \sin\theta'_1\sin\frac{\phi_1}{2} ~, \label{eq:suppl_newc2}\\
k\cos\theta'\sin\frac{\phi}{2} = \cos\frac{\phi_1}{2} ~.\label{eq:suppl_newc3}
\end{gather}
Here, we have two unknowns ($\theta'_1$ and $\phi_1$) and three equations. 
However, the three equations are not independent.
By squaring both sides of Eqs.~(\ref{eq:suppl_newc1}) and (\ref{eq:suppl_newc2}) and adding them, we obtain
\begin{equation}
 \cos^2\frac{\phi}{2} + \sin^2\theta' \sin^2\frac{\phi}{2} = \sin^2\frac{\phi_1}{2} ,
\end{equation}
which leads to
\begin{equation}
\cos^2\theta' \sin^2\frac{\phi}{2} = \cos^2\frac{\phi_1}{2} .
\end{equation}
This is the same as the square of Eq.~(\ref{eq:suppl_newc3}). 
We can obtain another equation by dividing Eq.~(\ref{eq:suppl_newc2}) by Eq.~(\ref{eq:suppl_newc1}):
\beq
k\sin\theta'\tan\frac{\phi}{2} = -\tan\theta'_1 ~. \label{eq:suppl_newc4}
\eeq

We now show that once $\theta'_1$ and $\phi_1$ are obtained, by solving Eqs.~(\ref{eq:suppl_newc3}) and (\ref{eq:suppl_newc4}), the results will also 
satisfy Eqs.~(\ref{eq:suppl_newc1}) and (\ref{eq:suppl_newc2}).
If we represent the left-hand-side (right-hand-side) of Eq.~(\ref{eq:suppl_newc1}) as $L_1$ ($R_1$), and similarly for $L_2$ ($R_2$) in Eq.~(\ref{eq:suppl_newc2}), then
Eq.~(\ref{eq:suppl_newc3}) implies that $L_1^2+L_2^2=R_1^2+R_2^2$, and Eq.~(\ref{eq:suppl_newc4}) leads to $L_2/L_1=R_2/R_1 \equiv \gamma$. 
Here we note that $L_2$ and $R_2$ are both positive, as was explained below Eq.~(\ref{eq:suppl_c2}). 
Plugging $L_1=L_2/\gamma$ and $R_1=R_2/\gamma$ into $L_1^2+L_2^2=R_1^2+R_2^2$, we obtain 
$ (1/\gamma^2+1) L_2^2 = (1/\gamma^2+1) R_2^2 $, and then $L_2^2=R_2^2$. 
Since $L_2$ and $R_2$ are positive, we obtain $L_2=R_2$.
Now, from  $L_2/L_1=R_2/R_1$, we obtain $L_1=R_1$. 
We can therefore determine $\phi_1$ from Eq.~(\ref{eq:suppl_newc3}) and $\theta'_1$ from Eq.~(\ref{eq:suppl_newc4}).
Note that Eqs.~(\ref{eq:suppl_newc3}) and (\ref{eq:suppl_newc4}) uniquely determine  $\phi_1$  in the range $[0,2\pi)$ and  $\theta'_1$ in the range $[0,\pi]$.

To summarize, we can always implement an arbitrary single-qubit gate with two rotation steps around axes in the $xz$ plane, given by $\phi_2=\pi$ and $ \theta'_2=0$, with $\phi_1$ obtained from Eq.~(\ref{eq:suppl_newc3}), and $\theta'_1$ obtained from Eq.~(\ref{eq:suppl_newc4}).
Of course, this is not the only solution. For example, we can also find a solution by choosing $\phi_1=\pi$. 
In that case $\theta'_1=0$, $e^{i\eta}=-k$, and after similar procedure we obtain
\begin{eqnarray}
-k\cos\theta'\sin\frac{\phi}{2} &=& \cos\frac{\phi_2}{2}~, \\
k\sin\theta'\tan\frac{\phi}{2} &=& \tan\theta'_2~, 
\end{eqnarray} 
which determine $\theta'_2$ and $\phi_2$ uniquely.
Once we determine $\theta'_1$ and $\theta'_2$, we can transform back to the original coordinate system using  $\theta_1$=$\theta'_1+\beta$ and $\theta_2$=$\theta'_2+\beta$.

\end{appendix}


\bibliographystyle{apsrev4-1}

\end{document}